\input ./mtexsis.tex
\texsis

\paper\tenpoint
\superrefsfalse 
\hfuzz 2.cm
\Eurostyletrue

\font\sstenz=cmss10 scaled \magstep0


\def\frac#1#2{{#1\over#2}}



\def\bra#1{\langle{#1}\vert}
\def\ket#1{\vert{#1}\rangle}




\def\n{\hfill\break}
\def\frac#1#2{{#1\over#2}}
%
%

\def\part{\partial}



\referencelist

\reference{[TGCVQT]} 
T.~Garavaglia, {\it Covariant relativistic quantum theory}, hep-th/9907059
\endreference


\reference{[HG]}
H.~Goldstein, {\it Classical Mechanics} 
(Addison Wesley, Reading Mass., 1959) p. 207
\endreference

\reference{[Fock]} 
V.~Fock, Physik Z. Sowjetunion {\bf 12}, 404 (1937) 
\endreference

\reference{[Schwinger]}
J.~Schwinger, Phys. Rev. {\bf 82}, 664 (1951)
\endreference

\reference{[Dirac33]} 
P.~A.~M.~Dirac, Physikalische Zeitschrift der Sowjetunion, Band {\bf 3},
Heft {\bf 1}, 64 (1933)
\endreference

\reference{[REDUCE]} 
A.~C.~Hearn, and J.~P.~Fitch, {\it REDUCE User's Manual}, (Rand, Santa Monica, 1995)
\endreference

\endreferencelist

\line{{cvefp.tex\hfill dias-stp-01-04}}
\n\n
\centerline{\bf Covariant Quantum Green's Function for an Accelerated Particle}
\n
\centerline{T. Garavaglia$^{*}$}
\n
\centerline{\it Institi\'uid \'Ard-l\'einn Bhaile \'Atha Cliath, Baile
\'Atha Cliath  4, \'Eire$^{\dagger}$}
\vskip 24pt
\par
\vbox{\sstenz 
\par
Covariant relativistic quantum theory is used to study the covariant 
Green's function, which can be used to determine 
the proper time evolved wave functions
that are solutions to the covariant Schr\"odinger type equation for a 
massive spin
zero particle. The concept of covariant action is used to obtain
the Green's function for an accelerated  relativistic particle. 
} 
\vskip 40pt
\noindent
PACS: {\sstenz 04.60, 03.65, 03.65P, 03.30}
\vskip 200pt 
\vskip 24pt
\hrule \n
\item{$^*$}{Email address: bronco@stp.dias.ie }  
\item{$^\dagger$}{Also {\it Institi\'uid Teicneola\'iochta Bhaile 
\'Atha Cliath.}}
\vfill\eject
\par
Covariant relativistic quantum theory has been described in \cite{[TGCVQT]}, and the
methods that are developed  there are used to find the proper time
covariant Green's function for a relativistic spin zero particle of mass $m$
that undergoes uniform acceleration. Covariant relativistic quantum
theory is a natural extension of the concepts of classical covariant 
relativistic particle 
dynamics \cite{[HG]}. Proper time covariant methods
have been  used previously in the context of relativistic wave equations
\cite{[Fock]}, and quantum field theory \cite{[Schwinger]}. 
The Lorentz invariant parameter that is relevant for dynamical development
in classical covariant mechanics is proper distance $s$, proper time times the
speed of light, and the invariant parameter that characterizes the rest
energy of a particle is its mass. 
The evolution of
quantum states with respect to the parameter $s$ is found in terms of 
the covariant Green's function. This Green's function is found with the
aid of the concept of covariant action. The resulting Green's function  has a practical
application in the study of the quantum state proper time evolution of, for 
example, an accelerated spin zero heavy ion.
\par
The conventions used are those of \cite{[TGCVQT]}, along with the units
$\hbar=c=1$.
The space-time contravariant coordinate four-vector is defined as
$q^\mu=(q^0,q^1,q^2,q^3)=(ct,x,y,z),$
and the dimensionless form is found by dividing the components with the
fundamental units of length $q_s$. The Einstein summation convention is used
throughout, and the covariant components are found from
$q_\mu=g_{\mu\nu}q^\nu$,
with the non-zero components of the metric tensor given by
$(g_{00},g_{11},g_{22},g_{33})=(1,-1,-1,-1).$
The scalar product of two four-vectors is $q\cdot p=q_\mu p^\mu$, and
the space-time measure is
$ds=\sqrt{dq\cdot dq}$.
The four-momentum for a free particle is
$p^\mu=mu^\mu=(p^0,\vec p)=(\gamma m,\gamma{\vec \beta}m)$,
with ${\vec \beta}={{d{\vec q}}/{dq^0}}$,
$\gamma=1/\sqrt{1-{\vec \beta}\cdot{\vec \beta}}$, $p\cdot p=m^2$, and
$u^\mu=dq^\mu/ds$.


\par
Covariant action is defined as
$$
{\cal S}=\int {\cal L}(q,u)ds,
\EQN{ac.1}$$
where ${\cal L}(q,u)$ is a Lorentz invariant Lagrangian.
The requirement that the action be an extremum under a variation of the
coordinates $q^\mu$ leads to the covariant Euler-Lagrange equation
$$
\frac{d \frac{\partial {\cal L}}{\partial u^\mu}}{ds}
-\frac{\partial {\cal L}}{\partial q^\mu}=0.
\EQN{ac.2}$$
\par
The covariant Lagrangian for a particle of mass $m$ and charge $e$ interacting 
with a four-vector field $A^\mu(q)$ is 
$$
{\cal L}=\frac{m}{2}u^2-eA\cdot u,
\EQN{ac.3}$$
and the generalized four-momentum is
$$
p_\mu=\frac{\partial {\cal L}}{\partial u^\mu}=mu_\mu-eA_\mu.
\EQN{ac.4}$$
The equation of motion is found from \Eq{ac.2} to be
$$
\frac{dp_\mu}{ds}+\partial_\mu{eA\cdot u}=0,
\EQN{ac.5}$$
where $\partial_\mu=\partial/\partial q^\mu$.
\par
From the covariant Lagrangian \Eq{ac.3}, the covariant Hamiltonian is defined 
as 
$$
{\cal H}=p\cdot u-{\cal L}=\frac{(p+eA)^2}{2m}.
\EQN{ac.6}$$


\par 
The properties of the covariant action can be used to determine the
covariant quantum Green's function for an accelerated spin zero 
particle of unit mass. For the four-vector potential $eA^\mu=(-V(\vec
q),0,0,0)$, the differential equation for the covariant action is
$$
\frac{1}{2}((\frac{\partial S}{\partial q^0}-
V(\vec q))^2-\vec \nabla S\cdot\vec
\nabla S)+\frac{\partial S}{\partial s}=0.
\EQN{qef.1}$$
The four-momentum associated with the covariant action is 
$$
p_\mu=\frac{\partial S}{\partial q^\mu}.
\EQN{qef.2}$$
For a free particle of unit mass  the solution to \Eq{qef.1}  is 
$$
S=\frac{(q-q')\cdot (q-q')}{2s}.
\EQN{qef.3}$$
\par
For a time independent potential $V(z)$, $p^0$ is a constant, and the covariant action can be written as
$$
S(q,s)=S_0(z,p^0,s)-\frac{y^2}{2s}-\frac{x^2}{2s}+p^0t;
\EQN{qef.4}$$
hence, \Eq{qef.1} becomes
$$
\frac{1}{2}((p^0-V(z))^2-(\frac{\partial S_0(z,p^0,s)}{\partial z})^2)
+\frac{\partial S_0(z,p^0,s)}{\partial s}=0.
\EQN{qef.5}$$
From \Eq{ac.4} and \Eq{ac.5}, the equations of motion become
$$\eqalign{
\frac{dt}{ds}&=p^0-V(z),\quad \frac{dp^0}{ds}=0,\cr
\frac{dz}{ds}&=p^3,\quad \frac{dp^3}{ds}=-(p^0-V(z))\frac{\partial V(z)}{\partial z}.\cr
}
\EQN{qef.6}$$
A particle with constant acceleration results from the potential $V(z)=-gz$,
where $g$ is a constant.
With the replacements
$q^\mu\rightarrow q^\mu/\sqrt{g}$, $p^\mu\rightarrow p^\mu\sqrt{g}$, and $s\rightarrow
s/g$, the equations of motion become
$$
\frac{dt}{ds}=p^0+z=\xi,\quad\frac{d^2\xi}{ds^2}=\xi.
\EQN{qef.7}$$
The solutions to these equations are
$$
\eqalign{
\xi(s)&=\xi(0)\cosh(s)+b\sinh(s)\cr
t(s)&=\xi(0)\sinh(s)+b(\cosh(s)-1)+t(0).\cr
}
\EQN{qef.8}$$
The action function $S_0(\xi,s)$ is now found from
$$
S_0(\xi,s)=\int_0^s{\cal L}_0 ds,
=-\int_0^s \frac{1}{2}((\frac{d\xi}{ds})^2+\xi^2)ds.
\EQN{qef.9}$$
After integration and the substitution
$$
b=(\xi-\xi(0)\cosh(s))/\sinh(s),
\EQN{qef.10}$$
one finds
$$
S_0(\xi,s)=- (1/2)((\xi^2+\xi(0)^2)\cosh(s)-2\xi\xi(0))/\sinh(s).
\EQN{qef.11}$$
\par
The covariant Schr\"odinger type equation for a spin zero particle is
$$
{\hat{\cal H}}\ket{\Psi(s)}=\frac{(\hat p+eA)^2}{2}\ket{\Psi(s)}
=i\frac{\partial \ket{\Psi(s)}}{\partial s},
\EQN{qef.12}$$
and the proper time evolved state is found from
$$
\ket{\Psi(s)}=\hat U(s)\ket{\Psi(0)}=e^{-i\hat{\cal H}s} \ket{\Psi(0)},
\EQN{qef.13}$$
The covariant proper time Green's function is
$$
{\cal G}(\vec q,q',s)=\bra{q}\hat U(s)\ket{q'}\theta(s),
\EQN{qef.14}$$
where $\theta(s)$ is the Heaviside function. This function is a solution to
the partial differential equation
$$
(\hat{\cal H}-i\frac{\partial}{\partial s}){\cal G}(q,q',s)=
-i\delta(q-q')\delta(s),
\EQN{qef.15}$$
and it satisfies the condition
$$
\lim_{s\rightarrow 0}{\cal G}(q,q',s)=\delta(q-q').
\EQN{qef.16}$$
The covariant Hamiltonian operator for the accelerated particle is 
$$
\hat{\cal H}=((\hat p^0+z)^2-\hat {\vec p}\cdot \hat {\vec p})/2,
\EQN{qef.17}$$
with $\hat p_\mu=i\partial/\partial q^\mu$, which satisfy the commutation
relation
$$
[\hat q^\mu,\hat p^\nu]=-ig^{\mu\nu}.
\EQN{qef.18}$$
\par
The s-dependent operators are
$$
\eqalign
{
\hat\xi(s)&=\hat U(s)\hat \xi \hat U(-s)=\hat\xi \cosh(s)+{\hat p^3} \sinh(s)\cr
\hat t(s)&=\hat U(s)\hat t \hat U(-s)=t+\hat\xi \sinh(s)+{\hat p^3}(\cosh(s)-1),\cr
}
\EQN{qef.19}$$
and the solutions to the classical equations of motion \Eq{qef.7} are found
from the matrix elements $\bra{\Psi(0)}\hat\xi(s)\ket{\Psi(0)}$, and 
$\bra{\Psi(0)}\hat t(s)\ket{\Psi(0)}$, where $\ket{\Psi(0)}$ is a minimum 
uncertainty state \cite{[TGCVQT]}.
Using the identities
$$\eqalign
{
\bra{q}\hat U(s)\hat \xi \hat U(-s)\hat U(s)\ket{q'}&=(\hat p^0+\hat z')\bra{q}\hat U(s)\ket{q'}\cr
\bra{q}\hat U(s)\hat t \hat U(-s)\hat U(s)\ket{q'}&=\hat t'\bra{q}\hat U(s)\ket{q'},\cr
}
\EQN{qef.20}$$
one can show that the Green's function also satisfies the first order
partial differential equations
$$
\eqalign
{
((\hat p^0+\hat z)\cosh(s)+\hat p^3\sinh(s)){\cal G}(q,q',s)=(\hat p^0+z'){\cal G}(q,q',s)\cr
(\hat t +(\hat p^0+\hat z)\sinh(s)+\hat p^3(\cosh(s)-1)){\cal G}(q,q',s)=t'{\cal G}(q,q',s).\cr
}
\EQN{qef.21}$$
\par
The solution to the equations \Eq{qef.15} and \Eq{qef.21} can be found using a method inspired by Dirac\cite{[Dirac33]} 
for non-relativistic quantum mechanics. One
assumes a solution of the form
$$
\bra{q}\hat U(s)\ket{q'}=a_0(s)\int_{-\infty}^{+\infty}e^{i\Phi}dp^0,
\EQN{qef.22}$$
with
$$
\Phi=S_0(z,z',p0,s)-\frac{(y-y')^2}{2s}
-\frac{(x-x')^2}{2s}-p^0(t-t'),
\EQN{qef.23}$$
and $q'^\mu=q^\mu(0)$.
The evaluation of \Eq{qef.22} and substitution into \Eq{qef.15}
leads to the differential equation for $a_0(s)$,
$$
\frac{da_0(s)}{ds}=-(1/s+\coth(s/2)/2)a_0(s).
\EQN{qef.24}$$
Solving this equation and using the condition \Eq{qef.16} gives the 
final result
$$
{\cal G}(q,q',s)=\frac{-i\theta(s)e^{i(-B^2A+C)}e^{-i\frac{(x-x')^2}{2s}}
e^{-i\frac{(y-y')^2}{2s}}}{8\pi^2s\sinh(s/2)},
\EQN{qef.25}$$
with
$$
\eqalign{
A(s)&=-\tanh(s/2)\cr
B(t,t',z,z',s)&=\frac{1}{2}(z+z'+(t-t')\coth(s/2))\cr
C(z,z',s)&=(-\cosh(s)\frac{1}{2}(z^2+z'^2)+zz')\sinh^{-1}(s).\cr
}
\EQN{qef.26}$$

The evolved wave function $\Psi(q,s)$ is found from the
initial state wave function $\Psi(q,0)$ using the evolution equation
$$
\Psi(q,s)=\int{\cal G}(q,q',s) \Psi( q',0)dq'.
\EQN{qef.27}$$
\par
Examples of evolved wave functions for a free particle are found in
\cite{[TGCVQT]}. In these examples the sum of the exponents of the initial
state wave function
and the Green's function are combined by completing the square,
and the $q'$ integration is then  performed. For a Gaussian initial state, 
it is
found that the evolved free particle wave function spreads about the 
classical path
described by the coordinates $q^\mu(s)=(p^0s,\vec p s)/m$, with $s=t/\gamma$.
For plane wave,
and Gaussian initial state wave functions, the integration using \Eq{qef.25}
can be performed in a similar manner.
Normal units may be restored in \Eq{qef.25} with the replacements
$q^\mu\rightarrow {q^\mu}/{q_s}$, $p^\mu\rightarrow {p^\mu}/{p_s}$,
and $s\rightarrow {s}/{s_s}$, with $q_s=\sqrt{\hbar c/(mg)}$, $q_sp_s=\hbar$, 
and $s_s=c^2/g$. In the limit $g\rightarrow 0$, the Green's function 
\Eq{qef.25} becomes the free particle
Green's function,
$$
{\cal G}(q,q',s)\rightarrow \frac{-i(mc)^2}{4(\pi\hbar s)^2}e^{i\frac{mc(q-q')^2}{2s\hbar}}.
\EQN{qef.28}$$
Calculations have been confirmed with the aid of REDUCE \cite{[REDUCE]}.
\n
\hrule
\n

\item{$^*$}{Email address: bronco@stp.dias.ie }

\item{$^\dagger$}{Also {\it Institi\'uid Teicneola\'iochta Bhaile \'Atha Cliath.}}

\ListReferences
\vfill\eject
\bye